# Quality related publication categories in social sciences and humanities, based on a university's peer review assessments

Nadine Rons*[1], Arlette De Bruyn[2]

[1] *Nadine.Rons@vub.ac.be*, [2] *Arlette.De.Bruyn@vub.ac.be*
Vrije Universiteit Brussel (VUB), B-1050 Brussels (Belgium)

**Introduction**

Bibliometric analysis has firmly conquered its place as an instrument for evaluation and international comparison of performance levels. Consequently, differences in coverage by standard bibliometric databases installed a dichotomy between on the one hand the well covered 'exact' sciences, and on the other hand most of the social sciences and humanities with a more limited coverage (Nederhof, 2006). Also the latter domains need to be able to soundly demonstrate their level of performance and claim or legitimate funding accordingly. An important part of the output volume in social sciences appears as books, book chapters and national literature (Hicks, 2004). To proceed from publication data to performance measurement, quantitative publication counts need to be combined with qualitative information, for example from peer assessment or validation (European Expert Group on Assessment of University-Based Research, 2010), to identify those categories that represent research quality as perceived by peers. An accurate focus is crucial in order to stimulate, recognize and reward high quality achievements only. This paper demonstrates how such a selection of publication categories can be based on correlations with peer judgments. It is also illustrated that the selection should be sufficiently precise, to avoid subcategories negatively correlated with peer judgments. The findings indicate that, also in social sciences and humanities, publications in journals with an international referee system are the most important category for evaluating quality. Book chapters with international referee system and contributions in international conference proceedings follow them.

**Method and material**

Ratings by peers and publication counts per full time equivalent leading staff (linked to promoter and funding opportunities) were collected from assessments per discipline by international expert panels at the Vrije Universiteit Brussel (Rons et al., 2008). The evaluations in social sciences and humanities involved 6 disciplines, 56 teams, near 500 full time equivalent researchers and 58 experts from 10 countries, and were conducted between 1999 and 2009. The 23 available publication categories span the total range from scientific publications to categories aimed at a professional and a broad audience. Categories that are only present for a minority of the teams in a discipline are not taken into account to avoid accidental occurrences. The 8 collected, interrelated peer rating categories are the overall evaluation score and scores on scientific merit, planning, innovation, team quality, feasibility, productivity and scientific impact. Correlations for the social sciences and humanities as a whole are calculated after normalization per discipline. The same methodology has been applied before to a different set of disciplines and also to other types of performance measures (Rons and De Bruyn, 2007).

## Table 1. Significant correlations with peer ratings per scientific publication category

| Medium:<br>Scope: | Book | Chap<br>Int | Jour<br>Int | Chap<br>Nat | Jour<br>Nat | Chap<br>no | Jour<br>no | Edit | Conf<br>Int | Abst<br>Int | Conf<br>Oth | Abst<br>Oth |
|---|---|---|---|---|---|---|---|---|---|---|---|---|
| **Social Sc. & Humanities** | | | 8+ | 4- | 7- | 1- | 6- | | | | | 1- |
| Psychology & Educat. Sc. | 2- | | 7+ | 2- | | | | | | | | |
| History | | | | | 3- | | 8- | 8- | | | | |
| Political & Social Sciences | 1+ | | 3+ | | | | | | | 1+ | | |
| Philosophy & Letters | 3- | 3+ | 5+ | | | 5- | 4- | | 5+ | 1- | | 6- |
| Economics | | | 5+ | 2- | 8- | | | | | | | |
| Law (particular categories) | 3+ | | 6+ | | | | | | | | | |

Number of peer rating categories (out of 8), per discipline, and for all social sciences and humanities disciplines combined, for which significantly positive or negative correlations are found with the publication category.
*Medium & scope:* **Book**s, book **Chap**ters and **Jour**nal articles with **Int**ernational, **Nat**ional or **no** referee system; **Edit**ed books or journals; Communications at **Int**ernational or **Oth**er conferences, integrally published (**Conf**) or published as abstract or not (**Abst**). Exception: particular publication categories for Law.

## Observations and conclusions

Table 1 highlights significantly positive and negative correlations with one or more peer rating categories at a 5% confidence level, for publications in books, journals and conference proceedings. Publication categories with an international dimension, in particular journal articles, show no other than positive correlations, while no other than negative or mixed correlations are found for the other categories. This indicates that in social sciences and humanities these 'international' publication categories can be used as legitimate general counterparts for the international journal publications focused on in exact sciences, with the intrinsically largely locally oriented discipline of Law as the exception to the rule. This also pinpoints the international dimension as an important criterion for selection or weighting of publication categories in performance based funding or evaluation systems, in order to stimulate quality as perceived by peers. In a context of best practices, it supports the rationale that, regardless of the discipline, high quality research performance requires that results be submitted to a sufficient extent to the scrutiny of the international research community. The particularly strong correlations with peer judgments found for the category of international journals suggest that this is the most effective publication medium for this purpose.

## References

Expert Group on Assessment of University-Based Research (2010). *Assessing Europe's University-Based Research*. DG Research, EUR 24187 EN.
Hicks, D. (2004). The Four Literatures of Social Science. In H.F. Moed, W. Glänzel and U. Schmoch (Eds.), *Handbook of Quantitative Science and Technology Research* (pp. 473-496). Kluwer Academic Publishers.
Nederhof, A.J. (2006). Bibliometric monitoring of research performance in the Social Sciences and the Humanities: A review. *Scientometrics* 66(1), 81-100.
Rons, N., De Bruyn, A. (2007). Quantitative CV-based indicators for research quality, validated by peer review. In D. Torres-Salinas and H. Moed (Eds.), *Proceedings of ISSI 2007,* 11th International Conference of the International Society for Scientometrics and Informetrics, CSIC, Madrid, Spain, 25-27 June 2007 (pp. 930-931).
Rons, N., De Bruyn, A., Cornelis, J. (2008). Research evaluation per discipline: a peer-review method and its outcomes. *Research Evaluation* 17(1), 45-57.